\documentclass[fleqn,10pt]{wlscirep}

\usepackage[utf8]{inputenc}
\usepackage[T1]{fontenc}
\usepackage{graphicx}
\usepackage{dcolumn}
\usepackage{bm}
\usepackage{gensymb}
\usepackage{siunitx}
\usepackage{chemformula}

\title{Ambient temperature pressure driven alkane dehydrogenation by palladium metal} 

\author[1,*]{Mungo Frost}
\author[1,2]{Emma E. McBride}
\author[3]{Dean Smith}
\author[3]{Jesse S. Smith}
\author[1]{Siegfried H. Glenzer}

\affil[1]{High Energy Density Science Division, SLAC National Accelerator Laboratory, 2575 Sand Hill Road, Menlo Park, USA}
\affil[2]{Stanford PULSE Institute, SLAC National Accelerator Laboratory, 2575 Sand Hill Road, Menlo Park, USA}
\affil[3]{High Pressure Collaborative Access Team, X-ray Science Division, Argonne National Laboratory, Argonne, USA}

\affil[*]{mdfrost@slac.stanford.edu}

\begin{abstract}
Dehydrogenation of alkanes is of increasing importance in fulfilling global demand for olefins and offers a potential source of carbon-neutral hydrogen as a co-product.  Currently commercial dehydrogenation processes occur at high-temperatures (500-900 \si{\celsius}) which is energy intensive and results in side reactions and rapid coking of the catalysts. In addition the hydrogen produced is often burned to maintain temperature and to inhibit the back reaction.  Here we demonstrate pressure as a parameter to enable novel chemical catalytic processes and demonstrate ambient-temperature dehydrogenation of alkanes by palladium at 50-100 MPa pressures, with both hydrogen gas and olefins recovered on decompression.  This reaction follows a fundamentally different path to current commercial high-temperature low-pressure dehydrogenation processes with the palladium catalyst reversibly forming a hydride intermediate. 
\end{abstract}

\begin{document}

\flushbottom
\maketitle

Hydrogen is of key importance in decarbonizing the global energy landscape as a fuel source for applications which are hard to electrify.  It has also gained attention as a possible means to store excess grid energy from renewables \cite{Peng22NatureEnergy}.  Various technologies are used to produce hydrogen, with its extraction from fossil fuel sources playing an important role, for example by methane pyrolysis \cite{Sanchez20}. These processes can avoid atmospheric carbon emissions by landfilling the solid carbon byproduct.  Alkane dehydrogenation is an alternative route to extract hydrogen from multi-carbon alkanes without \ch{CO2} emission. The alkanes can be sourced directly from oil, synthesized from \ch{CO2} \cite{Zhou22PNAS}, or made from methane by via non-oxidative coupling which itself produces hydrogen \cite{Zhang22NatureComms}.

Dehydrogenation of alkanes is a process by which hydrogen is removed from the alkane converting it to an olefin.  Alkanes are lower value hydrocarbons typically burned as fuels, while olefins are much more chemically versatile and are the feedstocks for many processes, such as the manufacture of plastics, other chemicals and pharmaceuticals.  As the carbon remains in the olefin product, the hydrogen is produced without release of \ch{CO2} and constitutes an environmentally conscious energy source.  Increased extraction of shale gas, compared to conventional oil, yields a greater portion of short chain alkanes which reduces the quantity of olefins available from processes such as cracking \cite{Sattler14}.  Coupled with increasing demand for olefins this leads to greater need for alkane dehydrogenation processes.  The overall reaction may be written as:

\ch{Alkane <> Olefin + Hydrogen}

Current industrial-scale dehydrogenation processes take place in the gas phase at high temperatures, 500 to 900\si{\celsius}, and low pressures, 20 - 300 kPa (0.2 to 3 bar), using solid chromia or platinum catalysts \cite{Sattler14, Perez12Book, Li21ChemSocRev}.  High temperatures are required for the favorable entropy from the loss of hydrogen to overcome the unfavorable enthalpy of the reaction (typically 28-30 kcal/mol).  The forward reaction is favored by low pressure as it converts the gaseous reactant to two molar equivalents of gaseous product.  High temperatures have several drawbacks such as poor selectivity and rapid coke formation which poisons the catalyst leading to the requirement for very frequent regeneration.  Side reactions such as cracking, and isomerization and metathesis of the olefin product are also more pronounced at higher temperature \cite{Sattler14, Perez12Book, Li21ChemSocRev, Liwu90AppCatalysis, Weckhuysen99CatalysisToday}.  The processes are energy intensive, and the hydrogen produced is frequently not recovered \cite{Sattler14}.

A large number of studies have investigated dehydrogenation using homogeneous catalysts \cite{Perez12Book, Li21ChemSocRev}.  These can have better selectivity and have the advantage of working at much lower temperatures.  However, they operate via hydrogen transfer, whereby the hydrogen removed from the alkane is transferred to a suitable acceptor molecule.  The choice of acceptor molecule can tune the enthalpy of the overall reaction to make it more favorable.  These all have the drawback of requiring an acceptor molecule and not allowing for recovery of molecular hydrogen.  The spent acceptor must also be considered for disposal or regeneration.

\begin{figure}
\centering
\includegraphics[width=0.7\columnwidth]{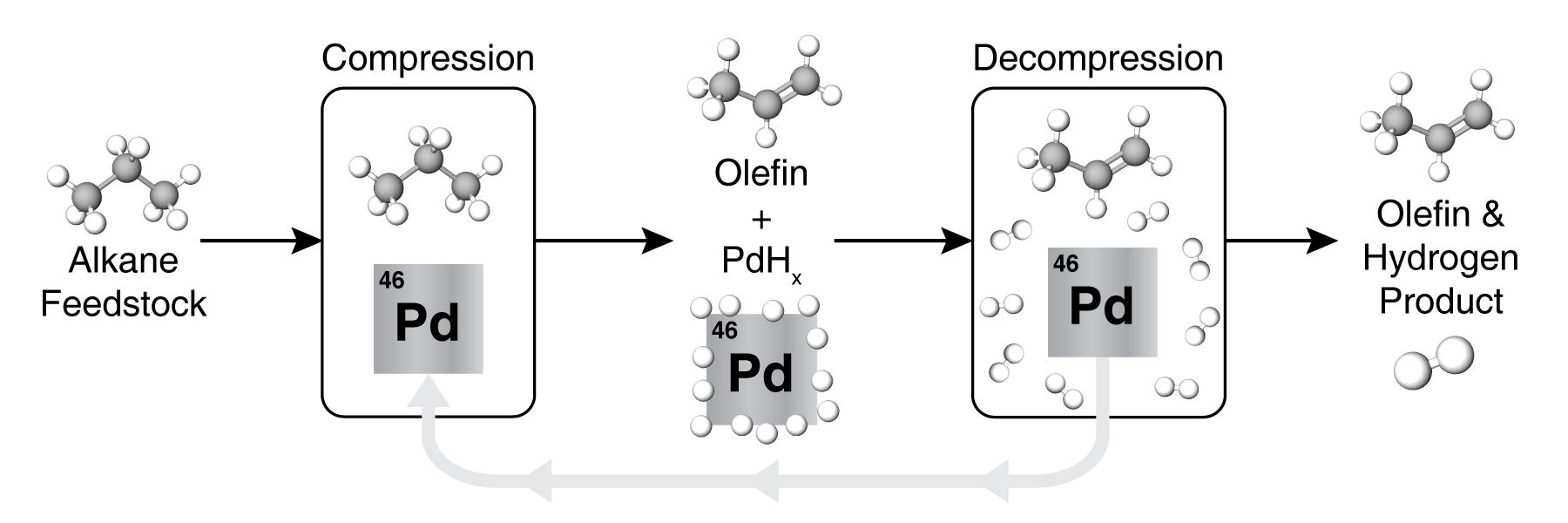}
\caption{Diagram of the palladium catalyzed dehydrogenation process.}
\label{fig:process}
\end{figure}

Here we report a novel route to the dehydrogenation of alkanes via a high-pressure intermediate.  Our process occurs in the condensed phase, and has the two key advantages in that it proceeds at ambient temperature and that the hydrogen is fully recovered.  The process has two steps. First the alkane is compressed in the presence of palladium metal.  This catalyzes dehydrogenation and acts as a hydrogen acceptor, forming palladium hydride as an intermediate, such that under pressure the alkane and palladium are converted to olefin and palladium hydride.  On pressure release the olefin persists and hydrogen is released from the palladium. In this study we show that the hydrogen gas is fully recoverable.  The overall reaction is: \ch{Alkane -> Olefin + Hydrogen}.  The carbon remains in the olefin for future use and the process releases no \ch{CO2}.  The process cycle is shown schematically in Figure \ref{fig:process}.

The high-pressure dehydrogenation is insensitive to the composition of the initial alkane.  We have observed it to occur in every alkane we have studied: n-tetracosane (\ch{C20H42}), n-octane (\ch{C8H18}), methylcyclohexane (\ch{C7H14}), and mineral oil (mixed alkanes).  This covers alkanes with 7 to 24 carbons, as well as mixtures, and we expect it to be applicable to any alkane except methane which cannot form a double bond as it has only one carbon.  To our knowledge, ambient-temperature, pressure-driven dehydrogenation has not been reported.  As well as offering an alternative to traditional alkane dehydrogenation, this process could produce clean hydrogen from low value feedstocks such as heavy fuel oils and bitumen as well as high weight byproducts from plastic thermolysis \cite{Alsalem09WasteManagement}.

We find the pressure required to be in the 50 to 100 MPa range.  It is worth noting that this magnitude of pressure is comfortably within the reach of industry and is commonly used in the fuel rails of diesel engines \cite{Xu18AppliedEnergy}, high pressure polyethylene synthesis \cite{Yoon85}, and food preservation \cite{Muntean16}, among others.

A key part of the process is the palladium hydride intermediate which forms on compression.  Palladium hydride has been widely studied \cite{Manchester94, Guigue20JAppPhys, Wise75JLessCommonMetals, Jamieson76JLessCommonMetals, Hatlevik10, Bugaev17JPhysChemC}.  There are two phases relevant to this room temperature study: the low-hydrogen $\alpha$ phase, and the high-hydrogen $\beta$ phase.  Both share the face centered cubic $Fm\bar{3}m$ structure of pure palladium, with the hydrogen occupying the octahedral interstices. They are nonstoichiometric, with more hydrogen resulting in a larger unit cell volume.  The $\alpha$ phase, \ch{PdH_x} with $x < 0.017$, is essentially palladium metal with trace dissolved hydrogen.  The $\beta$ phase is \ch{PdH_x}, $0.6 < x < 1$ and has much more hydrogen in the lattice.  At room temperature, hydrogen contents between 0.017 and 0.6 are forbidden, instead a mixture of $\alpha$ and $\beta$ phases form.  Higher hydrogen partial pressures lead to higher $x$ in the $\beta$ phase \cite{Manchester94, Guigue20JAppPhys}.  

At \SI{20}{\celsius}, the $\beta$ phase forms at around 3.5 kPa of hydrogen partial pressure, and $x$ saturates at 1 around 2 GPa \cite{Guigue20JAppPhys}.  At low hydrogen partial pressure, it will dissociate into the $\alpha$ phase and hydrogen gas.  This has led to it being proposed as a hydrogen storage system whereby hydrogen can be reversibly inserted into the palladium in a manner modulated by pressure \cite{Adams11MaterialsToday}.  At its core, our discovery shows that hydrogen can be extracted, stored, and released from palladium using alkanes as the hydrogen source rather than hydrogen gas; this process is accompanied by the simultaneous transformation of the alkane to an olefin.  Techniques developed to optimize hydrogen storage in palladium should also apply to the process described here.  These include controlling alloyants, temperature and particle size \cite{Hatlevik10, Manchester94, Yamauchi08JPhysChemC}.


\section*{Results}
\subsection*{Atomic Scale Insight from Diamond Anvil Cell Study}
The reaction was studied at the atomic scale using angle-resolved synchrotron powder x-ray diffraction in diamond anvil cells (DACs).  In these experiments the sample is held between two opposing diamond anvils and contained within a metallic gasket.  As the anvils are forced together the pressure on the sample increases.  DAC samples are microscopic, in this study between 100 and 200 $\mu$m diameter and 20 to 40 $\mu$m thick.

Samples consisted of powdered palladium embedded in the hydrocarbon being studied with a spherical ruby pressure marker \cite{Dewaele08PRB}.   Further details may be found in the Methods section.  Three hydrocarbon sources were investigated: methylcyclohexane, a relatively light alkane with formula \ch{C7H14}; n-tetracosane, a high molecular weight wax with formula \ch{C24H50}; and mineral oil which is a blend of several hydrocarbons.  These acted as both reactant and pressure transmitting medium and were in large chemical excess throughout.

\begin{figure}
\centering
\includegraphics[width=0.5\columnwidth]{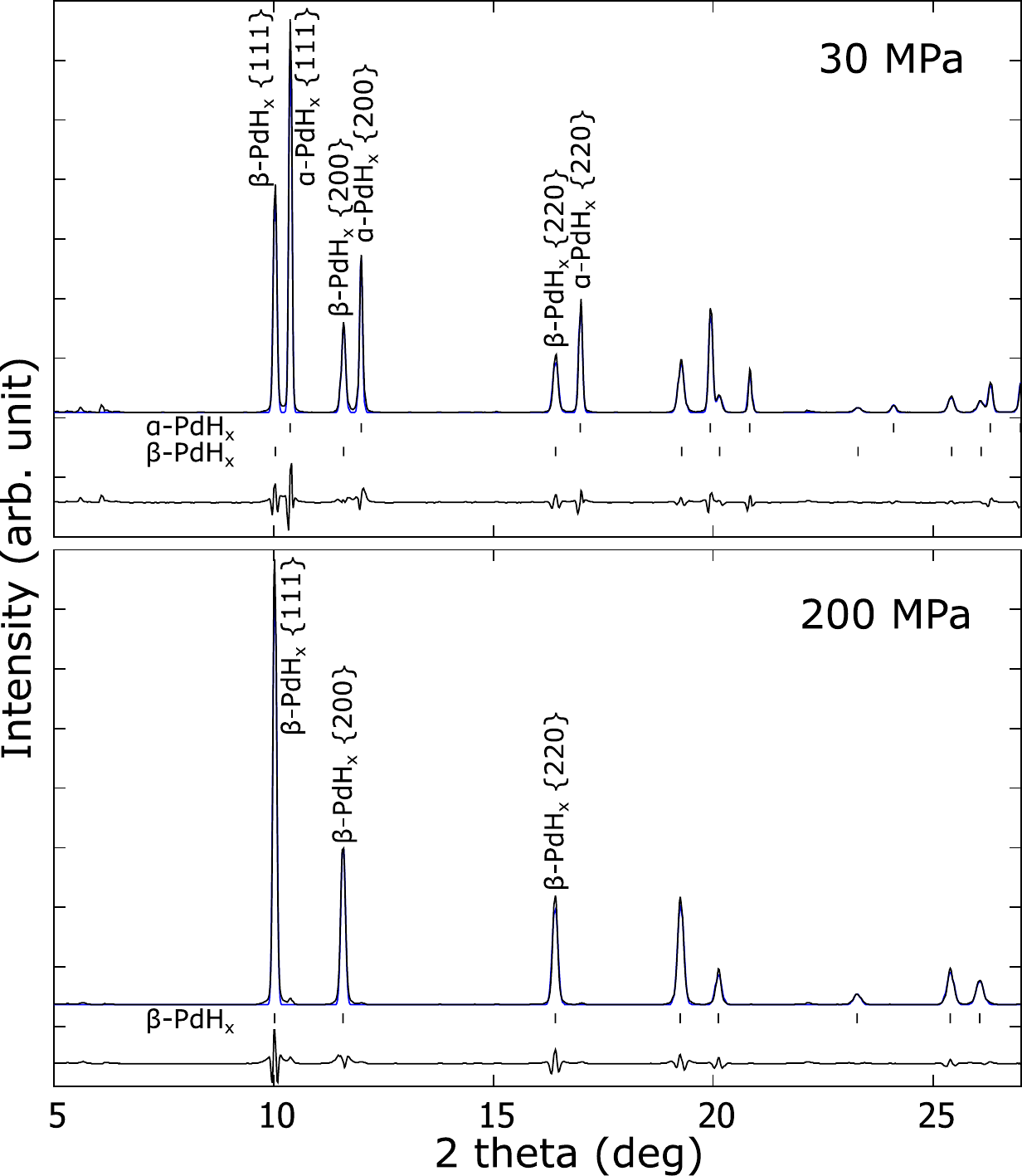}
\caption{Powder x-ray diffraction patterns of palladium hydrides formed from palladium in contact with n-tetracosane at high pressure.  Data is in black with LeBail fit in blue, the lower black line shows the residual. At 30 MPa (top) coexistence of the $\alpha$ and $\beta$ phases is observed, by 200 MPa (bottom) the $\alpha$ hydride is fully converted to the $\beta$. The diffraction patterns were collected using 30 keV radiation ($\lambda = 0.4066$ \AA).}
\label{fig:xrd}
\end{figure}

All the hydrocarbons studied underwent dehydrogenation.  This was indicated by initial splitting of the palladium peaks as the $\alpha$ hydride partially converted to the hydrogen-rich $\beta$ hydride before being fully consumed.  Integrated diffraction patterns are shown in Figure \ref{fig:xrd}.  Palladium was also compressed in a neon pressure transmitting medium with all sources of hydrogen strictly excluded, and the observed pressure-volume states closely matched a recently published equation of state for pure palladium \cite{Guigue20JAppPhys}.

\begin{figure}
\centering
\includegraphics[width=0.5\columnwidth]{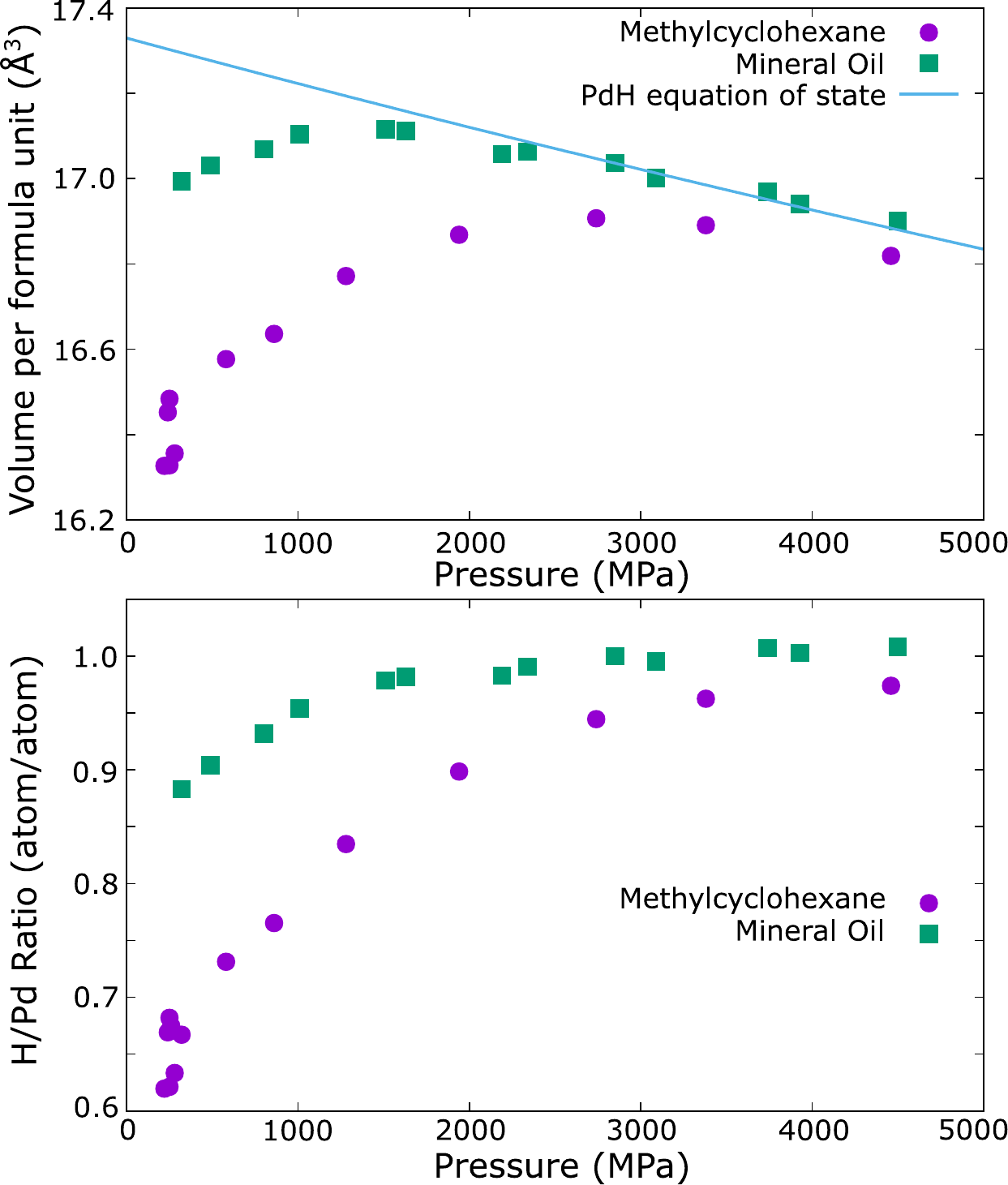}
\caption{\textbf{Upper:} Evolution of the unit cell volume of palladium hydride formed on compression in methylcyclohexane and mineral oil.  The equation of state of saturated palladium hydride, PdH$_{1.0}$, is shown \cite{Guigue20JAppPhys}.  \textbf{Lower:} Hydrogen loading of the palladium hydride with pressure calculated from the measured volumes and equations of state of Pd and PdH \cite{Guigue20JAppPhys}.  These observations indicate that the hydrogen content of palladium compressed at ambient temperature in alkanes is similar to that observed when it is compressed in molecular hydrogen.}
\label{fig:PV_Hloading}
\end{figure}

Further increases in pressure resulted in increased lattice volume, see Figure \ref{fig:PV_Hloading}.  Increasing volume with pressure is only possible if there is a change in composition, in this case additional hydrogen being absorbed into the lattice.  The palladium becomes saturated with hydrogen at 2 to 3 GPa, above which further compression reduces the lattice volume.  

Hydrogen saturated palladium hydride, \ch{PdH_x} where $x = 1$, can be produced by compression of palladium in hydrogen gas to 2 GPa \cite{Guigue20JAppPhys}.  The excess unit cell volume due to the hydrogen in the lattice can be calcaulated as a function of pressure by comparing the equations of state of the saturated hydride and pure palladium metal \cite{Guigue20JAppPhys}.  This allows allows for determination of the stoichiometry of the palladium hydride formed in this study by comparing the observed volume to that of pure palladium at the same pressure.  See Methods section for further discussion.  The hydrogen content of the hydride as a function of pressure is shown in Figure \ref{fig:PV_Hloading}.  The stoichiometry of the \ch{PdH_x} increases towards $x = 1$ with increasing pressure, as observed for palladium in pure hydrogen. 

\begin{figure}
\centering
\includegraphics[width=0.5\columnwidth]{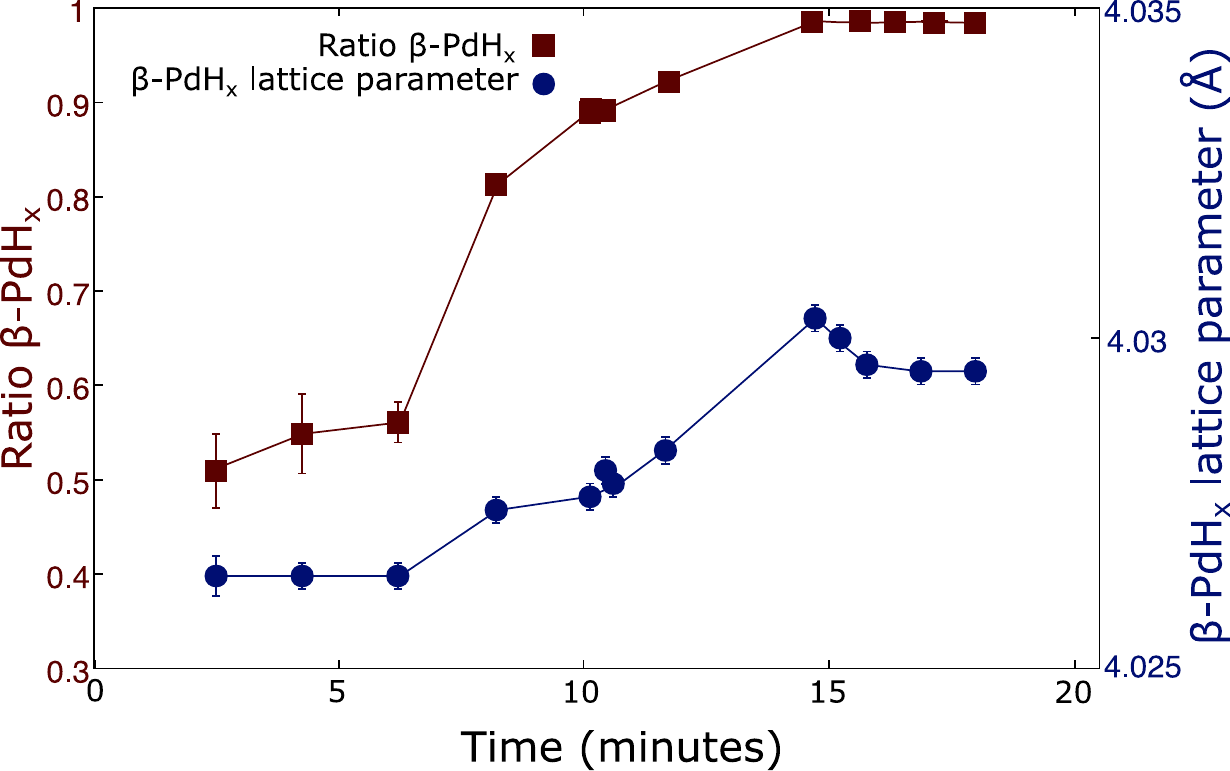}
\caption{Time resolved data on the ratio of $\beta$ to $\alpha$ palladium hydride (red), and the lattice parameter of the $\beta$ palladium hydride (blue).  Time is from pressure increase to 170 MPa.}
\label{fig:timeResolvedH}
\end{figure}

Time resolved powder x-ray diffraction data of hydride formation in palladium compressed in n-tetracosane on pressure increase to 170 MPa is shown in Figure \ref{fig:timeResolvedH}.  There is a short delay between compression and rapid hydride formation which is consistent with nucleation dynamics reported in other hydride systems \cite{Inomata98JAlloyComp, Vigeholm87JLessCommonMetals}, though in general such reactions are complex \cite{Bennett82PRB, Langhammer10PRL}.  After pressure increase there is a delay of approximately five minutes after which there is a rapid conversion of the residual $\alpha$ phase to the $\beta$ phase and slight increase in the $\beta$ lattice parameter is observed.  This delay is kinetic, arising from the energy cost of the initial formation of the $\beta$ hydride in the $\alpha$ as has been previously observed in the $\beta \rightarrow \alpha$ transition \cite{Jamieson76JLessCommonMetals}.  Care should be taken in the interpretation of the data as non-hydrostatic components of the stress in a DAC may be significant at such low pressure \cite{Takemura21HPR}.

\begin{figure}
\centering
\includegraphics[width=0.5\columnwidth]{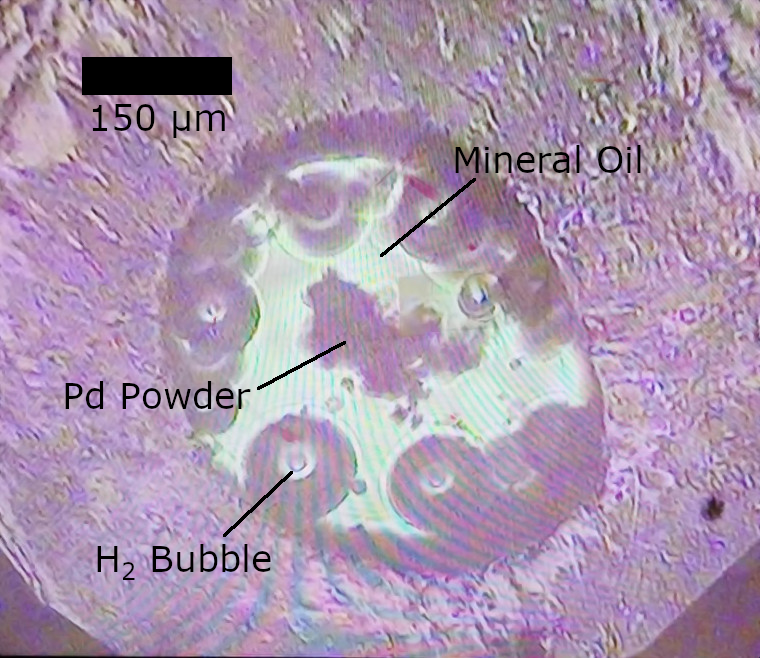}
\caption{Photomicrograph of a DAC sample of palladium in mineral oil after decompression.}
\label{fig:DACphoto}
\end{figure}

On decompression to ambient pressure gas evolution is observed from the palladium in DACs which forms bubbles in the liquid hydrocarbons.  A photomicrograph of a cell shortly after decompression is shown in Figure \ref{fig:DACphoto}.  This is compatible with hydrogen being released from the $\beta$ hydride as pressure is decreased.

\subsection*{Milliliter-Scale High-Pressure Reactor}

\begin{figure}
\centering
\includegraphics[width=0.5\columnwidth]{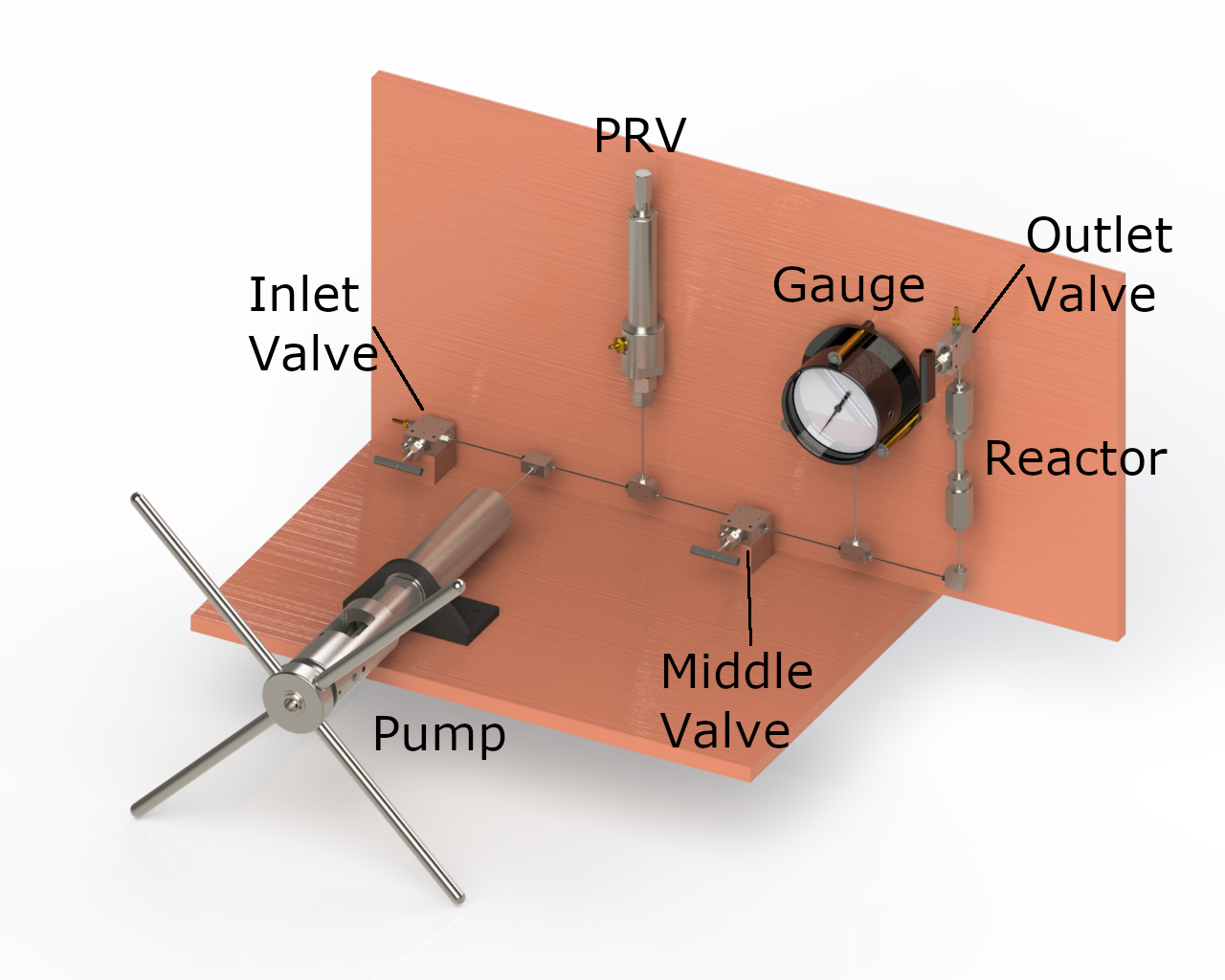}
\caption{A render of the high-pressure reactor is shown.  The system is 1 m long.  PRV is a pressure release valve safety to prevent over pressure.}
\label{fig:reactor}
\end{figure}

The reaction was studied at macroscopic volume in a high-pressure reactor with much finer control of pressure.  The system is shown in Figure \ref{fig:reactor}, pressure is generated by a pump connected to a 3 ml reaction chamber with a maximum working pressure of 400 MPa and pressure measurement and control of $\pm$2 MPa.  250 mg of powdered palladium was sealed into the reaction chamber and then the fluid of interest pumped into the desired pressure.  The system is large enough scale to allow samples of the olefin and hydrogen produced to be collected for further analysis.  

\begin{figure}
\centering
\includegraphics[width=0.7\columnwidth]{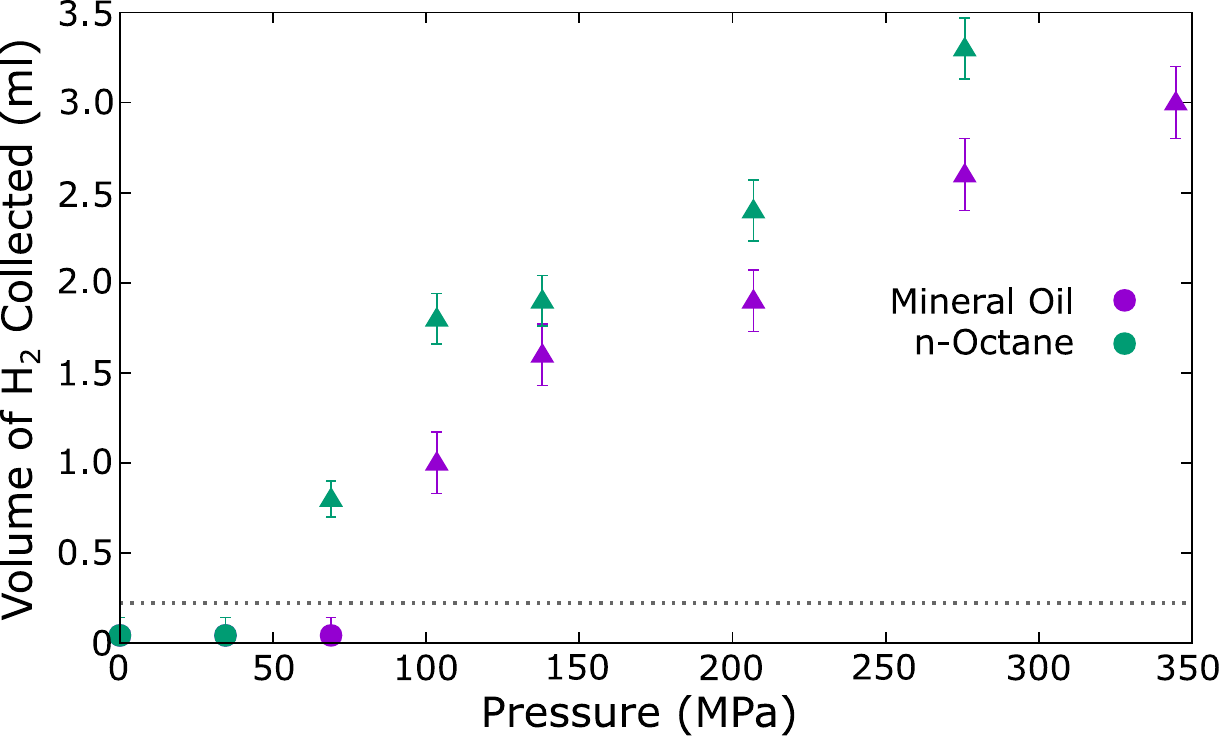}
\caption{Volume of hydrogen collected after compression of alkane precusors with 250 mg of palladium powder.  The dotted line at 0.224 ml is the maximum hydrogen possible without the formation of $\beta$ palladium hydride.}
\label{fig:reactorVvsP}
\end{figure}

The hydrocarbons tested were mineral oil, n-octane (\ch{C8H18}) and ethanol (\ch{C2H5OH}).  Both mineral oil and n-octane produced hydrogen on decompression, while ethanol exhibited no reaction up to 350 MPa, the maximum pressure studied.  A plot of the volume of hydrogen collected vs pressure for mineral oil and n-octane is shown in Figure \ref{fig:reactorVvsP}.  The gas collected was analyzed using an electrochemical hydrogen gas detector.  This confirmed the gas to be hydrogen, as expected from the diamond anvil cell measurements.

Figure \ref{fig:reactorVvsP} shows that hydrogen is produced from n-octane at 69 MPa (10 kpsi) and from mineral oil at 103 MPa (15 kpsi). Higher pressures lead to enhanced yield. Each reaction was held at pressure for 15 minutes: this is longer than the timescale required in the DAC experiment (Figure \ref{fig:timeResolvedH}) and should allow for reaction to occur.  Complete conversion of 250 mg of palladium metal to $\alpha$ palladium hydride with its maximum hydrogen stoichiometry of $x=0.017$ would allow for a maximum of 0.224 ml of hydrogen to be recovered.  More than this implies formation of the $\beta$ phase.  Complete conversion to the $\beta$ phase at its minimum hydrogen stoichiometry of $x=0.6$ would imply that there is 7.9 ml of hydrogen available for recovery, with a theoretical maximum yield of 13.2 ml for $x = 1$.

In this test study the yields were between these values.  Any recovery greater than 0.224 ml implies formation of the $\beta$ phase and the increased yields at higher pressure follow from higher hydrogen contents occurring, as is observed in the DAC experiment, see Figure \ref{fig:PV_Hloading}.  On pressure release hydrogen evolution from palladium hydride is known to occur in two steps \cite{Bennett82PRB}.  First the hydrogen content of the $\beta$ phase rapidly decreases to the lowest allowed value of PdH$_{0.6}$, after which there is a slower reaction as the $\beta$ hydride converts to the hydrogen-poor $\alpha$ hydride releasing further hydrogen gas.  Complete conversion to the $\alpha$ phase typically requires approximately 30 mbar hydrogen partial pressure \cite{Syrenova15NatureMat, Yamauchi08JPhysChemC}.  The very narrow capillaries in the high-pressure valves inhibit pulling a vacuum through them and prevent complete collection of hydrogen in this test system.  Full conversion back to palladium requires opening the reactor and putting the palladium catalyst under vacuum where the amount of hydrogen recovered is hard to quantify.  This limitation is due to the small scale of the test reactor and is not inherent to the process.

\begin{figure}
\centering
\includegraphics[width=0.7\columnwidth]{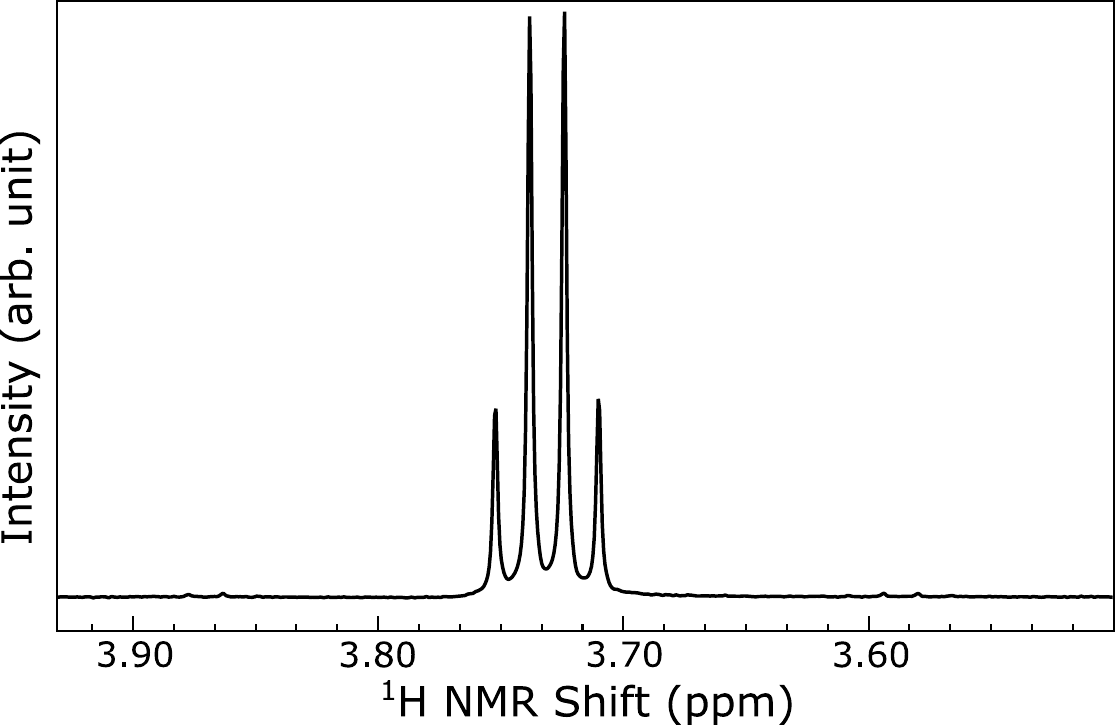}
\caption{Detail view of new peaks observed after compression of n-octane at 250 MPa in contact with palladium.  The NMR spectrum of the unreacted precursor n-octane has no detectable peaks in this region.  Full NMR spectra are shown in the supplementary materials.}
\label{fig:NMR}
\end{figure}

The liquid product was analyzed using proton nuclear magnetic resonance (NMR) spectroscopy.  Samples were recovered from n-octane compressed to 250 MPa in contact with palladium.  The n-octane precursor was chosen as, unlike mineral oil, it is a chemically pure species which simplifies interpretation of the NMR spectra.  In the test reactor the liquid being compressed acts both as the reactant and pressure transmitting fluid so the product is inherently dilute.  However, the recovered sample shows NMR peaks which correspond to the olefin product. Figure \ref{fig:NMR} shows the NMR spectrum of the recovered sample in the region from 3.5 to 3.9 ppm.  Vinylic protons, those connected to a carbon atom which is part of a carbon double bond, are expected in the 3.1 to 5.1 ppm region \cite{Atkins14Book}.  The reference NMR spectrum of the reactant exhibits no peaks in this region so we attribute these to the olefin product.  Further NMR spectra of both the unreacted n-octane and the product are shown in the supplementary materials.

\section*{Discussion}

Table \ref{table:summary} summarizes the substances investigated by different methods.  All alkanes and alkane mixtures exhibited reaction consistent with formation of $\beta$ palladium hydride at pressures around 100 MPa.  Ethanol, a primary alcohol, had no observable reaction.  One other study, performed at extreme pressure and high temperature, has reported palladium hydride formation from palladium and paraffin oil at 39(2) GPa and 1500(200) K \cite{Fedotenko20JAlloyComp}.  Such conditions are close to those required to convert hydrocarbons into diamond and hydrogen \cite{Benedetti99Science, Kraus17NatureAstro}, and metal hydrides are known to form from alkanes at such extreme reaction conditions in other systems \cite{Narygina11}.  That reaction is fundamentally different to the one reported here as it requires vastly higher pressure and temperatures to occur.  It is unclear why palladium hydride formation was not observed in ref \cite{Fedotenko20JAlloyComp} before laser heating, though it is possible that the heating was conducted prior to the synchrotron experiment and the unheated sample not measured. 

\begin{table}
\centering
\begin{tabular}{ | l c c c l | }
\hline
\textbf{Material} & \textbf{Formula} & \textbf{Dehydrogenation} & \textbf{Method} & \textbf{Comment} \\
\hline
n-Octane & \ch{C8H18} & Yes & High-Pressure Reactor & Alkane \\
n-Tetracosane & \ch{C24H50} & Yes & DAC & Alkane \\
Mineral Oil & Mixture & Yes & High-Pressure Reactor and DAC & Mixed Alkanes \\
Methylcyclohexane & \ch{C7H14} & Yes & DAC & Cycloalkane \\
Ethanol & \ch{C2H5OH} & No & High-Pressure Reactor & Alcohol \\
\hline
\end{tabular}
\caption{Summary of results.}
\label{table:summary}
\end{table}

The hydrogen content of palladium compressed with alkanes in this study, see Figure \ref{fig:PV_Hloading}, is similar to that of palladium compressed in pure hydrogen \cite{Guigue20JAppPhys}.  Both systems saturate at a 1:1 ratio of hydrogen:palladium at a few GPa which shows the dehydrogenation reaction to be favorable and is limited by the hydrogen affinity of the palladium.

In contrast to conventional heterogenous alkane dehydrogenation processes, which require high-temperatures and low-pressure, the process presented here occurs at ambient temperature and high-pressure.  This points to differences in the thermodynamics of the reactions.  All steps of the conventional reactions occur at the same conditions, and as it is overall endothermic and converts the gas phase reactant into two molar equivalents of gas phase products, the high-temperature and low-pressure reaction conditions are needed to drive the forward reaction.  Our process takes advantage of the pressure-tunable hydrogen affinity of palladium, as well as its catalytic properties, to force the forward reaction.  The reaction: \ch{alkane + palladium -> olefin + palladium~hydride} becomes increasingly favorable as pressure is increased. On pressure release the palladium hydride decomposes into palladium and hydrogen.  Further increase in the lattice parameter, and hence hydrogen content, of the palladium hydride on further pressure increase implies that both palladium metal and its hydrides catalyze this reaction.  In some respects the pressure driven dehydrogenation by palladium bears more similarities to homogenous catalytic dehydrogenation reactions \cite{Perez12Book, Li21ChemSocRev}, but with the palladium metal acting as both catalyst and a reversible hydrogen acceptor.

The kinetics of metal hydride formation and decomposition are complex and can depend on the purity, grain size, composition, and history of the material \cite{Syrenova15NatureMat, Yamauchi08JPhysChemC, Wise75JLessCommonMetals, Zadorozhnyy17JAlloyComp}.  Palladium and its hydrides have undergone significant study in this regard due to its uses and potential applications in hydrogen storage and purification, and in fuel cells.  Alloyants, temperature, sample morphology and particle size can all be used to tailor the kinetics of the reactions \cite{Syrenova15NatureMat, Yamauchi08JPhysChemC, Jamieson76JLessCommonMetals, Auer74}.  Palladium coating has also been used to enhance other hydrogen storage materials \cite{Barcelo10, Krozer90JLessCommonMetals}.  These approaches for optimization of hydrogen uptake by palladium might be extended to enhance the method described here.

It is interesting to note that the equation of state of palladium is the subject of some controversy in the literature.  Recently it has been measured twice, once to 62 GPa \cite{Guigue20JAppPhys} and once to 31 GPa \cite{Fedotenko20JAlloyComp}.  The equations of state for palladium reported are in disagreement, particularly when extrapolated to higher pressures.  The higher pressure study \cite{Guigue20JAppPhys} agrees well with our x-ray diffraction results on pure palladium in neon in a cell kept scrupulously clean of any source of hydrogen contamination.  It is likely that some of the other results, for example the extremely high pressure derivative of bulk modulus reported in ref. \cite{Fedotenko20JAlloyComp}, arise from low-level hydrogen contamination of the palladium sample owing to the previously unknown reaction presented here.  Such a reaction can occur with trace contaminants, for example grease on a loading needle or culet, and slightly enlarge a unit cell in a way which only fully presents itself after compression has begun.

In conclusion, we present a novel dehydrogenation reaction of alkanes which converts palladium metal to palladium hydride at pressures around 100 MPa.  The hydrogen may then be recovered from the palladium metal by decompression making this an alternate route to alkane dehydrogenation which has various advantages over conventional processes including removing the necessity of high-temperature and allowing full recovery of the hydrogen produced without the release of carbon dioxide to the atmosphere.  The reaction proceeded in all alkanes studied, but not in ethanol, a primary alcohol.  This reaction may also explain the controversy in the equation of state of palladium reported by prior studies.


\section*{Methods}

\subsection*{Diamond Anvil Cells}

Angularly-resolved synchrotron powder x-ray diffraction was carried out in DACs.  These were equipped with wide angle Bohler-Almax anvils with culets between 250 and 400 $\mu$m and rhenium gaskets preindented to 20 to 40 $\mu$m.  Pressure was determined via the fluorescence wavelength of a small ruby sphere included in each loading \cite{Dewaele08PRB}.

Samples were prepared several days prior to the experiment. The methylcyclohexane and mineral oil were loaded to 100 MPa, while the solid n-tetracosane was loaded to 30 MPa. Powder x-ray diffraction was performed at HP-CAT beamline 16-ID-B at the Advanced Photon Source using monochromatic 30 keV radiation with data collected on a Dectris Pilatus 1M large area detector equiped with \SI{172}{\micro\meter} pixels at a distance of 207.06 mm from the sample. Compression was performed stepwise using a gas membrane. On initially fitting the cells to the gas membrane apparatus there is a small (<30 MPa) increase in pressure, thus the initial measurements are made after a small pressure increase from the loaded pressures.

Data was integrated using the Dioptas software package \cite{Prescher15HPR}.  LeBail fits of the integrated data were performed using Jana2006 \cite{Petvrivcek14}, while the ratio of the hydride phases was estimated based on the relative intensities of the {111}, {200} and {220} diffraction peaks of each phase. As the structures are very similar and hydrogen contributes very little to the scattering this gives a good measure of the relative quantities of each phase in the patterns.

The dehydrogenation reaction occurs around 100 MPa.  DACs are typically used at much higher pressures and control and measurement of pressure in the very low pressure regime here is challenging.  Pressure changes of less than 100 MPa are difficult to achieve and pressure measurement based on ruby fluorescence is subject to 30 MPa uncertainty.  Nonetheless they are an invaluable tool for observing the reactants \textit{in situ} at high pressure and following the palladium content of the palladium hydride based on its lattice parameter determined by powder x-ray diffraction.  They also allowed the reaction to be observed to much higher pressures of several GPa.

\subsection*{Determination of Stoichiometry of Palladium Hydride}

The stoichiometry of the palladium hydride formed can be determined from the unit cell volume obtained from powder x-ray diffraction data.  The presence of hydrogen increases the unit cell volume.  The excess volume due to the hydrogen partially occupying the octahedral interstices of the lattice is simply the difference between the unit cell volume of the hydride and that of pure palladium metal.  This volume is then compared to the volume increase from a hydrogen atom sitting in the palladium lattice to give an average occupancy.  This method is commonly close to ambient pressure \cite{Fukai06Book}.  To account for the effect of pressure the known equations of state \cite{Guigue20JAppPhys} of palladium and saturated palladium hydride with a 1:1 ratio of Pd:H are compared.  This allows the volume of a hydrogen atom in the palladium lattice to be determined as a function of pressure, which can in turn be compared the the excess volume of the hydride under study.  Mathematically this may be expressed as x in \ch{PdH_x} being given by:

\begin{equation*}
x = \frac{V_{hydride}(P) - V_{Pd}(P)}{V_{PdH}(P) - V_{Pd}(P)}
\end{equation*}

Where $V_{hydride}$ is the measured volume of the \ch{PdH_x} at pressure $P$, $V_{Pd}$ is the volume of pure palladium and $V_{PdH}$ is the volume of the saturated palladium hydride with $x=1$.  $V_{Pd}$ and $V_{sat. hydride}$ can be calculated using a Vinet equation of state and the parameters from ref. \cite{Guigue20JAppPhys}. 

\subsection*{High-Pressure Test Reactor}

A high-pressure reactor allowed the reaction to be studied at macroscopic volume and with much finer control of pressure.  It is shown in Figure \ref{fig:reactor}, and consists of a pumped system with a 3 ml reaction chamber into which 250 mg of powdered palladium was sealed and then the fluid of interest pumped in.  Pressure is measured using a bourdon gauge and can be set and measured to 2 MPa precision up to a maximum pressure of 400 MPa.

The high-pressure reactor is of sufficient volume to allow hydrogen to be collected for further analysis.  It does not, however, allow for \textit{in situ} measurement as the vessel is constructed from stainless steel.  Complete recovery of the hydrogen is also challenging as both the hydrogen gas produced and residual hydrocarbon must traverse tight passages in the high-pressure valves which prohibit complete recovery.  Full removal of the hydrogen requires opening of the reactor so the palladium can be placed directly under vacuum.  These limitations of the test reactor could be easily overcome in a larger-scale custom designed reactor.

The gas collected was analyzed using an electrochemical hydrogen detector with NIST traceable calibration (Forensics Detectors model FD-90A-H2).  This measures from 0 to 1000 ppm hydrogen concentration with 5\% uncertainty.  The detector was placed in an airtight container of known volume, then a known volume of the collected gas inserted.  The resultant concentration of hydrogen gas in the container is then measured and confirms the gas collected to be hydrogen.

\subsection*{Proton Nuclear Magnetic Resonance}

Proton NMR was performed by an independent third party laboratory (NuMegaLab, San Diego, California) using 500 MHz spectrometers.  Samples were dissolved in \ch{CDCl3} for analysis and consisted of the liquid recovered after compression of n-octane to 250 MPa in contact with palladium in the high-pressure test reactor and the unreacted n-octane feedstock as a reference.  

\section*{Acknowledgements}
This work was supported by U.S. Department of Energy (DOE) Office of Fusion Energy Sciences funding No. FWP100182.  This work was supported by the Department of Energy, Laboratory Directed Research and Development program at the SLAC National Accelerator Laboratory under Contract No. DE-AC02-76SF00515 and as part of the Panofsky Fellowship awarded to E.E.M.  X-ray diffraction was performed at HPCAT (Sector 16), Advanced Photon Source (APS), Argonne National Laboratory. HPCAT operations are supported by DOE-NNSA’s Office of Experimental Sciences.  The Advanced Photon Source is a U.S. Department of Energy (DOE) Office of Science User Facility operated for the DOE Office of Science by Argonne National Laboratory under Contract No. DE-AC02-06CH11357.

\section*{Data Availability}
The datasets used and/or analysed during the current study available from the corresponding author on reasonable request.

\section*{Author contributions statement}
M.F. and S.H.G. conceived the experiment, M.F., E.E.M., D.S. and J.S.S. conducted the DAC experiments. M.F. designed, constructed and conducted experiments with the high-pressure reactor.  M.F. analyzed the results and wrote the manuscript.  All authors reviewed the manuscript. 

\section*{Competing Interests}
The pressure driven alkane dehydrogenation process described is under pending patient application by Stanford University under the names of Mungo Frost, Emma E. McBride, Dean Smith, Jesse S. Smith and Siegfried Glenzer.  Application number 63/389220, status at time of writing: Provisional US Patent.

\bibliography{Dehydrogenation}

\begin{thebibliography}{10}
\expandafter\ifx\csname url\endcsname\relax
  \def\url#1{\texttt{#1}}\fi
\expandafter\ifx\csname urlprefix\endcsname\relax\def\urlprefix{URL }\fi
\providecommand{\bibinfo}[2]{#2}
\providecommand{\eprint}[2][]{\url{#2}}

\bibitem{Peng22NatureEnergy}
\bibinfo{author}{Peng, P.} \emph{et~al.}
\newblock \bibinfo{title}{Cost and potential of metal--organic frameworks for
  hydrogen back-up power supply}.
\newblock \emph{\bibinfo{journal}{Nature Energy}} \bibinfo{pages}{1--11}
  (\bibinfo{year}{2022}).

\bibitem{Sanchez20}
\bibinfo{author}{S{\'a}nchez-Bastardo, N.}, \bibinfo{author}{Schl{\"o}gl, R.}
  \& \bibinfo{author}{Ruland, H.}
\newblock \bibinfo{title}{Methane pyrolysis for co2-free h2 production: A green
  process to overcome renewable energies unsteadiness}.
\newblock \emph{\bibinfo{journal}{Chemie Ingenieur Technik}}
  \textbf{\bibinfo{volume}{92}}, \bibinfo{pages}{1596--1609}
  (\bibinfo{year}{2020}).

\bibitem{Zhou22PNAS}
\bibinfo{author}{Zhou, C.} \emph{et~al.}
\newblock \bibinfo{title}{Steering co2 hydrogenation toward c--c coupling to
  hydrocarbons using porous organic polymer/metal interfaces}.
\newblock \emph{\bibinfo{journal}{Proceedings of the National Academy of
  Sciences}} \textbf{\bibinfo{volume}{119}}, \bibinfo{pages}{e2114768119}
  (\bibinfo{year}{2022}).

\bibitem{Zhang22NatureComms}
\bibinfo{author}{Zhang, W.} \emph{et~al.}
\newblock \bibinfo{title}{High-performance photocatalytic nonoxidative
  conversion of methane to ethane and hydrogen by heteroatoms-engineered tio2}.
\newblock \emph{\bibinfo{journal}{Nature Communications}}
  \textbf{\bibinfo{volume}{13}}, \bibinfo{pages}{1--9} (\bibinfo{year}{2022}).

\bibitem{Sattler14}
\bibinfo{author}{Sattler, J.~J.}, \bibinfo{author}{Ruiz-Martinez, J.},
  \bibinfo{author}{Santillan-Jimenez, E.} \& \bibinfo{author}{Weckhuysen,
  B.~M.}
\newblock \bibinfo{title}{Catalytic dehydrogenation of light alkanes on metals
  and metal oxides}.
\newblock \emph{\bibinfo{journal}{Chemical reviews}}
  \textbf{\bibinfo{volume}{114}}, \bibinfo{pages}{10613--10653}
  (\bibinfo{year}{2014}).

\bibitem{Perez12Book}
\bibinfo{author}{P{\'e}rez, P.~J.}
\newblock \emph{\bibinfo{title}{Alkane CH activation by single-site metal
  catalysis}}, vol.~\bibinfo{volume}{38} (\bibinfo{publisher}{Springer Science
  \& Business Media}, \bibinfo{year}{2012}).

\bibitem{Li21ChemSocRev}
\bibinfo{author}{Li, C.} \& \bibinfo{author}{Wang, G.}
\newblock \bibinfo{title}{Dehydrogenation of light alkanes to mono-olefins}.
\newblock \emph{\bibinfo{journal}{Chemical Society Reviews}}
  \textbf{\bibinfo{volume}{50}}, \bibinfo{pages}{4359--4381}
  (\bibinfo{year}{2021}).

\bibitem{Liwu90AppCatalysis}
\bibinfo{author}{Liwu, L.}, \bibinfo{author}{Tao, Z.},
  \bibinfo{author}{Jingling, Z.} \& \bibinfo{author}{Zhusheng, X.}
\newblock \bibinfo{title}{Dynamic process of carbon deposition on pt and pt--sn
  catalysts for alkane dehydrogenation}.
\newblock \emph{\bibinfo{journal}{Applied catalysis}}
  \textbf{\bibinfo{volume}{67}}, \bibinfo{pages}{11--23}
  (\bibinfo{year}{1990}).

\bibitem{Weckhuysen99CatalysisToday}
\bibinfo{author}{Weckhuysen, B.~M.} \& \bibinfo{author}{Schoonheydt, R.~A.}
\newblock \bibinfo{title}{Alkane dehydrogenation over supported chromium oxide
  catalysts}.
\newblock \emph{\bibinfo{journal}{Catalysis today}}
  \textbf{\bibinfo{volume}{51}}, \bibinfo{pages}{223--232}
  (\bibinfo{year}{1999}).

\bibitem{Alsalem09WasteManagement}
\bibinfo{author}{Al-Salem, S.}, \bibinfo{author}{Lettieri, P.} \&
  \bibinfo{author}{Baeyens, J.}
\newblock \bibinfo{title}{Recycling and recovery routes of plastic solid waste
  (psw): A review}.
\newblock \emph{\bibinfo{journal}{Waste management}}
  \textbf{\bibinfo{volume}{29}}, \bibinfo{pages}{2625--2643}
  (\bibinfo{year}{2009}).

\bibitem{Xu18AppliedEnergy}
\bibinfo{author}{Xu, L.} \emph{et~al.}
\newblock \bibinfo{title}{Experimental and modeling study of liquid fuel
  injection and combustion in diesel engines with a common rail injection
  system}.
\newblock \emph{\bibinfo{journal}{Applied energy}}
  \textbf{\bibinfo{volume}{230}}, \bibinfo{pages}{287--304}
  (\bibinfo{year}{2018}).

\bibitem{Yoon85}
\bibinfo{author}{Yoon, B.~J.} \& \bibinfo{author}{Rhee, H.-K.}
\newblock \bibinfo{title}{A study of the high pressure polyethylene tubular
  reactor}.
\newblock \emph{\bibinfo{journal}{Chemical Engineering Communications}}
  \textbf{\bibinfo{volume}{34}}, \bibinfo{pages}{253--265}
  (\bibinfo{year}{1985}).

\bibitem{Muntean16}
\bibinfo{author}{Muntean, M.-V.} \emph{et~al.}
\newblock \bibinfo{title}{High pressure processing in food
  industry--characteristics and applications}.
\newblock \emph{\bibinfo{journal}{Agriculture and Agricultural Science
  Procedia}} \textbf{\bibinfo{volume}{10}}, \bibinfo{pages}{377--383}
  (\bibinfo{year}{2016}).

\bibitem{Manchester94}
\bibinfo{author}{Manchester, F.}, \bibinfo{author}{San-Martin, A.} \&
  \bibinfo{author}{Pitre, J.}
\newblock \bibinfo{title}{The h-pd (hydrogen-palladium) system}.
\newblock \emph{\bibinfo{journal}{Journal of phase equilibria}}
  \textbf{\bibinfo{volume}{15}}, \bibinfo{pages}{62--83}
  (\bibinfo{year}{1994}).

\bibitem{Guigue20JAppPhys}
\bibinfo{author}{Guigue, B.}, \bibinfo{author}{Geneste, G.},
  \bibinfo{author}{Leridon, B.} \& \bibinfo{author}{Loubeyre, P.}
\newblock \bibinfo{title}{An x-ray study of palladium hydrides up to 100 gpa:
  Synthesis and isotopic effects}.
\newblock \emph{\bibinfo{journal}{Journal of Applied Physics}}
  \textbf{\bibinfo{volume}{127}}, \bibinfo{pages}{075901}
  (\bibinfo{year}{2020}).

\bibitem{Wise75JLessCommonMetals}
\bibinfo{author}{Wise, M.}, \bibinfo{author}{Farr, J.} \&
  \bibinfo{author}{Harris, I.}
\newblock \bibinfo{title}{X-ray studies of the $\alpha$/$\beta$ miscibility
  gaps of some palladium solid solution-hydrogen systems}.
\newblock \emph{\bibinfo{journal}{Journal of the Less Common Metals}}
  \textbf{\bibinfo{volume}{41}}, \bibinfo{pages}{115--127}
  (\bibinfo{year}{1975}).

\bibitem{Jamieson76JLessCommonMetals}
\bibinfo{author}{Jamieson, H.}, \bibinfo{author}{Weatherly, G.} \&
  \bibinfo{author}{Manchester, F.}
\newblock \bibinfo{title}{The $\beta$→ $\alpha$ phase transformation in
  palladium-hydrogen alloys}.
\newblock \emph{\bibinfo{journal}{Journal of the Less Common Metals}}
  \textbf{\bibinfo{volume}{50}}, \bibinfo{pages}{85--102}
  (\bibinfo{year}{1976}).

\bibitem{Hatlevik10}
\bibinfo{author}{Hatlevik, {\O}.} \emph{et~al.}
\newblock \bibinfo{title}{Palladium and palladium alloy membranes for hydrogen
  separation and production: History, fabrication strategies, and current
  performance}.
\newblock \emph{\bibinfo{journal}{Separation and Purification Technology}}
  \textbf{\bibinfo{volume}{73}}, \bibinfo{pages}{59--64}
  (\bibinfo{year}{2010}).

\bibitem{Bugaev17JPhysChemC}
\bibinfo{author}{Bugaev, A.~L.} \emph{et~al.}
\newblock \bibinfo{title}{Core--shell structure of palladium hydride
  nanoparticles revealed by combined x-ray absorption spectroscopy and x-ray
  diffraction}.
\newblock \emph{\bibinfo{journal}{The Journal of Physical Chemistry C}}
  \textbf{\bibinfo{volume}{121}}, \bibinfo{pages}{18202--18213}
  (\bibinfo{year}{2017}).

\bibitem{Adams11MaterialsToday}
\bibinfo{author}{Adams, B.~D.} \& \bibinfo{author}{Chen, A.}
\newblock \bibinfo{title}{The role of palladium in a hydrogen economy}.
\newblock \emph{\bibinfo{journal}{Materials today}}
  \textbf{\bibinfo{volume}{14}}, \bibinfo{pages}{282--289}
  (\bibinfo{year}{2011}).

\bibitem{Yamauchi08JPhysChemC}
\bibinfo{author}{Yamauchi, M.}, \bibinfo{author}{Ikeda, R.},
  \bibinfo{author}{Kitagawa, H.} \& \bibinfo{author}{Takata, M.}
\newblock \bibinfo{title}{Nanosize effects on hydrogen storage in palladium}.
\newblock \emph{\bibinfo{journal}{The Journal of Physical Chemistry C}}
  \textbf{\bibinfo{volume}{112}}, \bibinfo{pages}{3294--3299}
  (\bibinfo{year}{2008}).

\bibitem{Dewaele08PRB}
\bibinfo{author}{Dewaele, A.}, \bibinfo{author}{Torrent, M.},
  \bibinfo{author}{Loubeyre, P.} \& \bibinfo{author}{Mezouar, M.}
\newblock \bibinfo{title}{Compression curves of transition metals in the mbar
  range: Experiments and projector augmented-wave calculations}.
\newblock \emph{\bibinfo{journal}{Physical Review B}}
  \textbf{\bibinfo{volume}{78}}, \bibinfo{pages}{104102}
  (\bibinfo{year}{2008}).

\bibitem{Inomata98JAlloyComp}
\bibinfo{author}{Inomata, A.}, \bibinfo{author}{Aoki, H.} \&
  \bibinfo{author}{Miura, T.}
\newblock \bibinfo{title}{Measurement and modelling of hydriding and
  dehydriding kinetics}.
\newblock \emph{\bibinfo{journal}{Journal of Alloys and Compounds}}
  \textbf{\bibinfo{volume}{278}}, \bibinfo{pages}{103--109}
  (\bibinfo{year}{1998}).

\bibitem{Vigeholm87JLessCommonMetals}
\bibinfo{author}{Vigeholm, .~B.}, \bibinfo{author}{Jensen, K.},
  \bibinfo{author}{Larsen, B.} \& \bibinfo{author}{Pedersen, A.~S.}
\newblock \bibinfo{title}{Elements of hydride formation mechanisms in nearly
  spherical magnesium powder particles}.
\newblock \emph{\bibinfo{journal}{Journal of the Less Common Metals}}
  \textbf{\bibinfo{volume}{131}}, \bibinfo{pages}{133--141}
  (\bibinfo{year}{1987}).

\bibitem{Bennett82PRB}
\bibinfo{author}{Bennett, P.} \& \bibinfo{author}{Fuggle, J.}
\newblock \bibinfo{title}{Electronic structure and surface kinetics of
  palladium hydride studied with x-ray photoelectron spectroscopy and
  electron-energy-loss spectroscopy}.
\newblock \emph{\bibinfo{journal}{Physical Review B}}
  \textbf{\bibinfo{volume}{26}}, \bibinfo{pages}{6030} (\bibinfo{year}{1982}).

\bibitem{Langhammer10PRL}
\bibinfo{author}{Langhammer, C.}, \bibinfo{author}{Zhdanov, V.~P.},
  \bibinfo{author}{Zori{\'c}, I.} \& \bibinfo{author}{Kasemo, B.}
\newblock \bibinfo{title}{Size-dependent kinetics of hydriding and dehydriding
  of pd nanoparticles}.
\newblock \emph{\bibinfo{journal}{Physical review letters}}
  \textbf{\bibinfo{volume}{104}}, \bibinfo{pages}{135502}
  (\bibinfo{year}{2010}).

\bibitem{Takemura21HPR}
\bibinfo{author}{Takemura, K.}
\newblock \bibinfo{title}{Hydrostaticity in high pressure experiments: some
  general observations and guidelines for high pressure experimenters}.
\newblock \emph{\bibinfo{journal}{High Pressure Research}}
  \textbf{\bibinfo{volume}{41}}, \bibinfo{pages}{155--174}
  (\bibinfo{year}{2021}).

\bibitem{Syrenova15NatureMat}
\bibinfo{author}{Syrenova, S.} \emph{et~al.}
\newblock \bibinfo{title}{Hydride formation thermodynamics and hysteresis in
  individual pd nanocrystals with different size and shape}.
\newblock \emph{\bibinfo{journal}{Nature materials}}
  \textbf{\bibinfo{volume}{14}}, \bibinfo{pages}{1236--1244}
  (\bibinfo{year}{2015}).

\bibitem{Atkins14Book}
\bibinfo{author}{Atkins, P.}, \bibinfo{author}{Atkins, P.~W.} \&
  \bibinfo{author}{de~Paula, J.}
\newblock \emph{\bibinfo{title}{Atkins' physical chemistry}}
  (\bibinfo{publisher}{Oxford university press}, \bibinfo{year}{2014}).

\bibitem{Fedotenko20JAlloyComp}
\bibinfo{author}{Fedotenko, T.} \emph{et~al.}
\newblock \bibinfo{title}{Synthesis of palladium carbides and palladium hydride
  in laser heated diamond anvil cells}.
\newblock \emph{\bibinfo{journal}{Journal of Alloys and Compounds}}
  \textbf{\bibinfo{volume}{844}}, \bibinfo{pages}{156179}
  (\bibinfo{year}{2020}).

\bibitem{Benedetti99Science}
\bibinfo{author}{Benedetti, L.~R.} \emph{et~al.}
\newblock \bibinfo{title}{Dissociation of ch4 at high pressures and
  temperatures: diamond formation in giant planet interiors?}
\newblock \emph{\bibinfo{journal}{Science}} \textbf{\bibinfo{volume}{286}},
  \bibinfo{pages}{100--102} (\bibinfo{year}{1999}).

\bibitem{Kraus17NatureAstro}
\bibinfo{author}{Kraus, D.} \emph{et~al.}
\newblock \bibinfo{title}{Formation of diamonds in laser-compressed
  hydrocarbons at planetary interior conditions}.
\newblock \emph{\bibinfo{journal}{Nature Astronomy}}
  \textbf{\bibinfo{volume}{1}}, \bibinfo{pages}{606--611}
  (\bibinfo{year}{2017}).

\bibitem{Narygina11}
\bibinfo{author}{Narygina, O.} \emph{et~al.}
\newblock \bibinfo{title}{X-ray diffraction and m{\"o}ssbauer spectroscopy
  study of fcc iron hydride feh at high pressures and implications for the
  composition of the earth's core}.
\newblock \emph{\bibinfo{journal}{Earth and Planetary Science Letters}}
  \textbf{\bibinfo{volume}{307}}, \bibinfo{pages}{409--414}
  (\bibinfo{year}{2011}).

\bibitem{Zadorozhnyy17JAlloyComp}
\bibinfo{author}{Zadorozhnyy, V.~Y.} \emph{et~al.}
\newblock \bibinfo{title}{Effect of mechanical activation on compactibility of
  metal hydride materials}.
\newblock \emph{\bibinfo{journal}{Journal of Alloys and Compounds}}
  \textbf{\bibinfo{volume}{707}}, \bibinfo{pages}{214--219}
  (\bibinfo{year}{2017}).

\bibitem{Auer74}
\bibinfo{author}{Auer, W.} \& \bibinfo{author}{Grabke, H.}
\newblock \bibinfo{title}{The kinetics of hydrogen absorption in palladium
  ($\alpha$-and $\beta$-phase) and palladium-silver-alloys}.
\newblock \emph{\bibinfo{journal}{Berichte der Bunsengesellschaft f{\"u}r
  physikalische Chemie}} \textbf{\bibinfo{volume}{78}}, \bibinfo{pages}{58--67}
  (\bibinfo{year}{1974}).

\bibitem{Barcelo10}
\bibinfo{author}{Barcelo, S.}, \bibinfo{author}{Rogers, M.},
  \bibinfo{author}{Grigoropoulos, C.~P.} \& \bibinfo{author}{Mao, S.~S.}
\newblock \bibinfo{title}{Hydrogen storage property of sandwiched magnesium
  hydride nanoparticle thin film}.
\newblock \emph{\bibinfo{journal}{international journal of hydrogen energy}}
  \textbf{\bibinfo{volume}{35}}, \bibinfo{pages}{7232--7235}
  (\bibinfo{year}{2010}).

\bibitem{Krozer90JLessCommonMetals}
\bibinfo{author}{Krozer, A.} \& \bibinfo{author}{Kasemo, B.}
\newblock \bibinfo{title}{Hydrogen uptake by pd-coated mg:
  absorption-decomposition isotherms and uptake kinetics}.
\newblock \emph{\bibinfo{journal}{Journal of the Less Common Metals}}
  \textbf{\bibinfo{volume}{160}}, \bibinfo{pages}{323--342}
  (\bibinfo{year}{1990}).

\bibitem{Prescher15HPR}
\bibinfo{author}{Prescher, C.} \& \bibinfo{author}{Prakapenka, V.~B.}
\newblock \bibinfo{title}{Dioptas: a program for reduction of two-dimensional
  x-ray diffraction data and data exploration}.
\newblock \emph{\bibinfo{journal}{High Pressure Research}}
  \textbf{\bibinfo{volume}{35}}, \bibinfo{pages}{223--230}
  (\bibinfo{year}{2015}).

\bibitem{Petvrivcek14}
\bibinfo{author}{Pet{\v{r}}{\'\i}{\v{c}}ek, V.}, \bibinfo{author}{Du{\v{s}}ek,
  M.} \& \bibinfo{author}{Palatinus, L.}
\newblock \bibinfo{title}{Crystallographic computing system jana2006: general
  features}.
\newblock \emph{\bibinfo{journal}{Zeitschrift f{\"u}r
  Kristallographie-Crystalline Materials}} \textbf{\bibinfo{volume}{229}},
  \bibinfo{pages}{345--352} (\bibinfo{year}{2014}).

\bibitem{Fukai06Book}
\bibinfo{author}{Fukai, Y.}
\newblock \emph{\bibinfo{title}{The metal-hydrogen system: basic bulk
  properties}}, vol.~\bibinfo{volume}{21} (\bibinfo{publisher}{Springer Science
  \& Business Media}, \bibinfo{year}{2006}).

\end{thebibliography}

\end{document}


\preprint{}

\title{Ambient temperature pressure driven alkane dehydrogenation by palladium metal -- Supplementary Materials} 

\author{Mungo Frost}
\affiliation{%
 SLAC National Accelerator Laboratory, 2575 Sand Hill Road, Menlo Park, USA
}%

\author{Emma E. McBride}
 \affiliation{%
 SLAC National Accelerator Laboratory, 2575 Sand Hill Road, Menlo Park, USA
}%

\author{Dean Smith}
 \affiliation{%
  High Pressure Collaborative Access Team, X-ray Science Division, Argonne National Laboratory, Argonne, USA
}%

\author{Jesse S. Smith}
 \affiliation{%
  High Pressure Collaborative Access Team, X-ray Science Division, Argonne National Laboratory, Argonne, USA
}%

\author{Siegfried H. Glenzer}
\affiliation{%
 SLAC National Accelerator Laboratory, 2575 Sand Hill Road, Menlo Park, USA 
}%

\date{\today}

\maketitle

\section{Proton Nuclear Magnetic Resonance Spectra}

Proton NMR was performed by an independent third party laboratory (NuMegaLab, San Diego, California) using 500 MHz spectrometers.  Samples were dissolved in \ch{CDCl3} for analysis and consisted of the liquid recovered after compression of n-octane to 250 MPa in contact with palladium in the high-pressure test reactor and the unreacted n-octane feedstock as a reference.

\subsection{n-Octane Feedstock}

\begin{figure}[h]
\includegraphics[width=1\columnwidth]{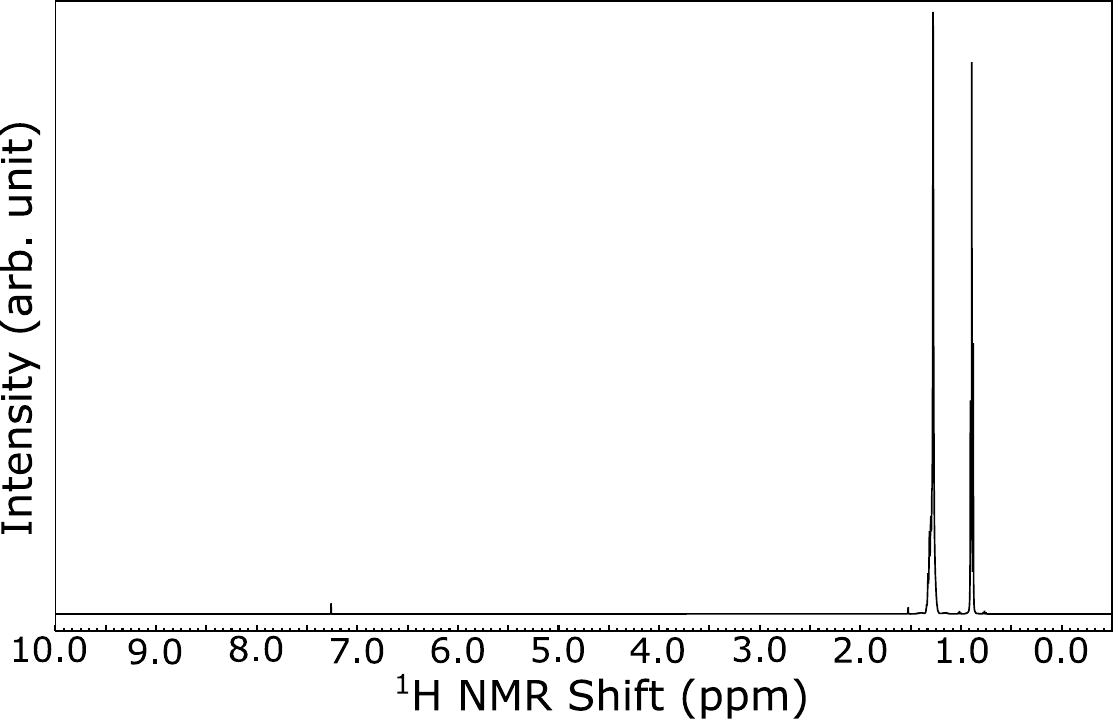}
\caption{NMR spectrum of the unreacted n-octane source material.}
\label{fig:noctaneNMR}
\end{figure}

The proton NMR spectrum of the n-alkane reactant is shown in Figure \ref{fig:noctaneNMR}.  It exhibits peaks around 0.89 and 1.30 ppm corresponding to \ch{CH3} and \ch{CH2} groups respectively.  The peak at 7.26 ppm arises from residual protons in the \ch{CDCl3} solvent.  The weak peak at 1.53 ppm is a trace contaminant, probably water.  

\subsection{Recovered Sample}

\begin{figure}[h]
\includegraphics[width=1\columnwidth]{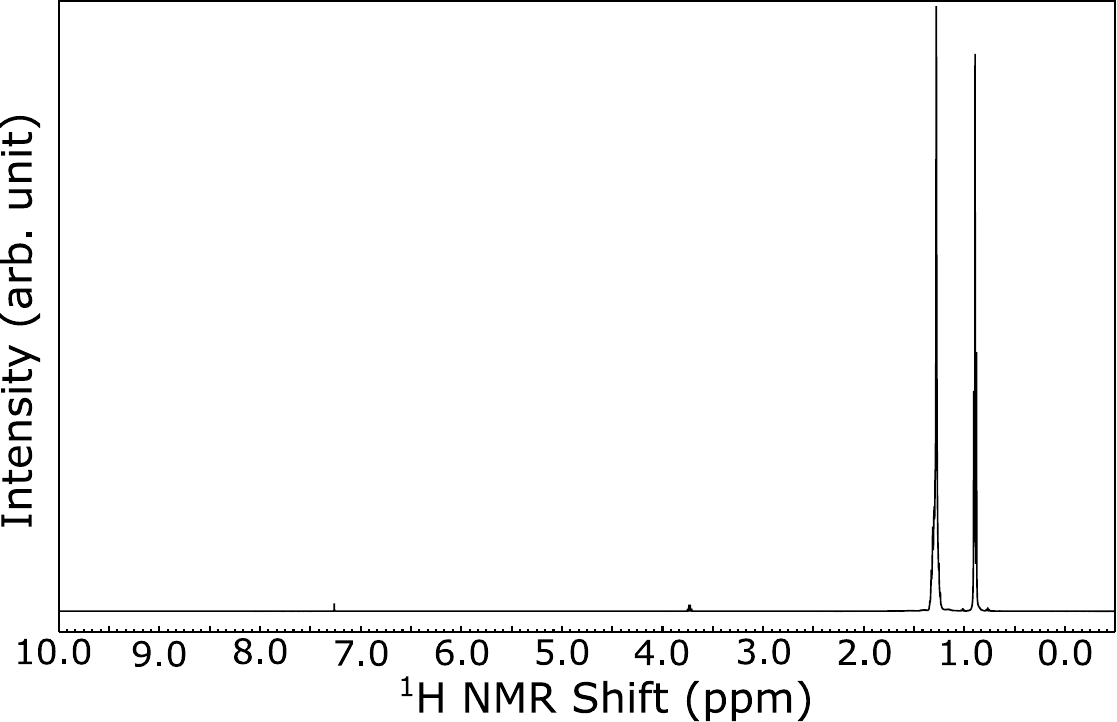}
\caption{NMR spectrum of the product after compression of n-octane in contact with palladium powder at 250 MPa.  Note the new peaks close to 3.73 ppm.}
\label{fig:productNMR}
\end{figure}

Figure \ref{fig:productNMR} shows an NMR spectrum of the recovered product.  There is a large excess of n-octane as the high-pressure test reactor uses it as both reactant and pressure medium leading to an inherently dilute product. However, new peaks are clearly visible around 3.73 ppm.  These are compatible with hydrogens bonded to carbon atoms which form part of a carbon-carbon double bond, consistent with an olefin product.   

\begin{figure}[h]
\includegraphics[width=1\columnwidth]{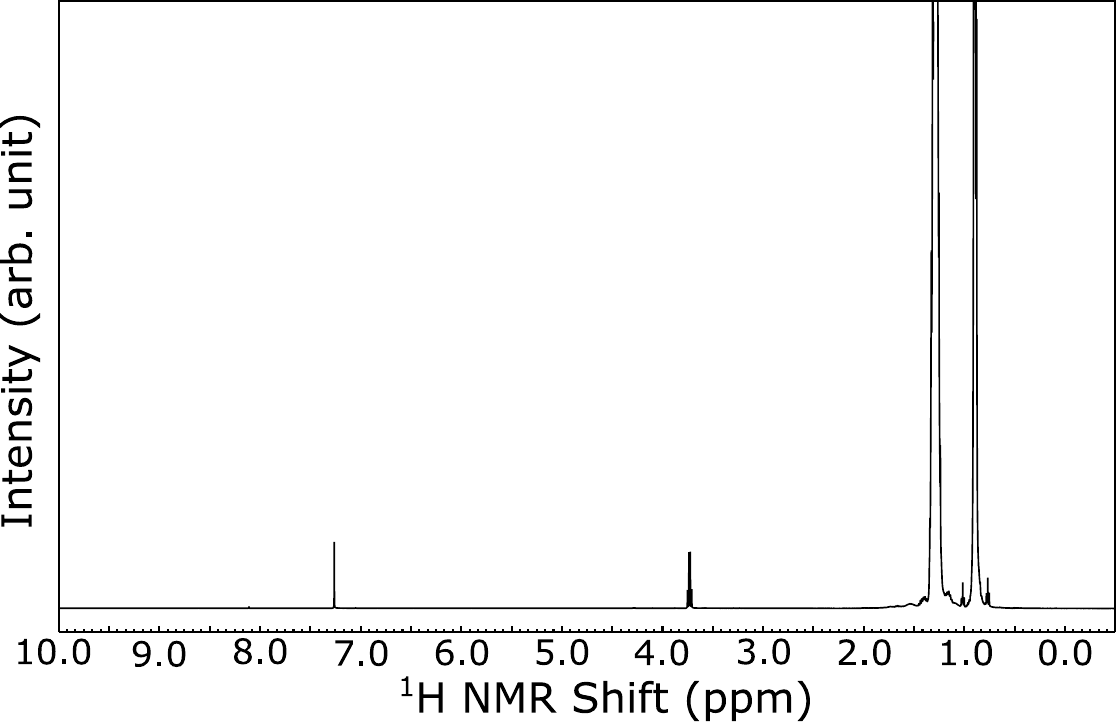}
\caption{Same spectrum as Figure \ref{fig:productNMR} with expanded y-scale.  A detailed view of the new peaks is shown in the main manuscript.}
\label{fig:productNMRzoom}
\end{figure}

Figure \ref{fig:productNMRzoom} shows this spectrum with the y-scale expanded to make weak peaks easier to see.

